\documentclass{article}

\usepackage{arxiv}

\usepackage[utf8]{inputenc} 
\usepackage[T1]{fontenc}    
\usepackage{hyperref}       
\usepackage{url}            
\usepackage{booktabs}       
\usepackage{amsfonts}       
\usepackage{nicefrac}       
\usepackage{microtype}      
\usepackage{lipsum}
\usepackage{graphicx}

\usepackage{array}

\usepackage{xltabular}

\usepackage{ragged2e}
\usepackage{xcolor}

\usepackage{longtable}

\usepackage[most]{tcolorbox}
\usepackage[table]{xcolor}
\usepackage{fontawesome5}

\date{}
\renewcommand{\today}{}
\AtBeginDocument{%
\pagestyle{fancy}%
\fancyhf{}
\fancyfoot[C]{\thepage}
}

\newcolumntype{P}[1]{>{\RaggedRight\arraybackslash}p{#1}}
\arrayrulecolor{black!30}          

\graphicspath{ {./images/} }

\tcbset{
boxC/.style={
colback=orange!5!white,
colframe=orange!20!white,
coltitle=orange!60!black,
fonttitle=\bfseries,
boxrule=1mm,
arc=4mm,
left=3mm,
right=3mm,
top=3mm,
bottom=3mm,
title={#1},
before skip=10pt,
after skip=10pt,
boxsep=5pt
},
}
\NewDocumentEnvironment{boxC}{O{} +b}{
    \begin{tcolorbox}[boxC,title={#1}]  
        #2
    \end{tcolorbox}
}{}

\title{Visual Authority and the Rhetoric of Health Misinformation: A Multimodal Analysis of Social Media Videos}

\author{
 Mohammad Reza Zarei \\
  School of Computer Science\\
  Carleton University\\
  \texttt{mohammadrezazarei@cmail.carleton.ca} \\
   \And
 Barbara Stead-Coyle \\
  Department of Law and Legal Studies\\
  Carleton University\\
  \texttt{Barbarasteadcoyle@cmail.carleton.ca} \\
  \And
 Michael Christensen \\
  Department of Law and Legal Studies\\
  Carleton University\\
  \texttt{Michael.Christensen@carleton.ca} \\
 \And
 Sarah Everts \\
  School of Journalism and Communication\\
  Carleton University\\
  \texttt{Sarah.Everts@carleton.ca} \\
 \And
 Majid Komeili \\
  School of Computer Science\\
  Carleton University\\
  \texttt{Majid.Komeili@carleton.ca} \\
} 

\begin{document}
\maketitle
\begin{abstract}
Short form video platforms are central sites for health advice, where alternative narratives mix useful, misleading, and harmful content. Rather than adjudicating truth, this study examines how credibility is packaged in nutrition and supplement videos by analyzing the intersection of authority signals, narrative techniques, and monetization. We assemble a cross platform corpus of 152 public videos from TikTok, Instagram, and YouTube and annotate each on 26 features spanning visual authority, presenter attributes, narrative strategies, and engagement cues. A transparent annotation pipeline integrates automatic speech recognition, principled frame selection, and a multimodal model, with human verification on a stratified subsample showing strong agreement. Descriptively, a confident single presenter in studio or home settings dominates, and clinical contexts are rare. Analytically, authority cues such as titles, slides and charts, and certificates frequently occur with persuasive elements including jargon, references, fear or urgency, critiques of mainstream medicine, and conspiracies, and with monetization including sales links and calls to subscribe. References and science like visuals often travel with emotive and oppositional narratives rather than signaling restraint.

\end{abstract}


\section{Introduction}
\label{sec:intro}
People increasingly turn to social media for personalized health advice \cite{Chen_Wang_2021}. During and after the COVID‑19 pandemic, alternative health entrepreneurs, political pundits, and influencers built large audiences with content that challenged public health guidance and cast doubt on health authorities \cite{Lundy_2023,Sommer_2023, Heley_Gaysynsky_King_2022, Brennen_Simon_Nielsen_2021}. Short-form video platforms such as TikTok, YouTube, and Instagram have become central sites for health information. Alongside credible advice, they also amplify alternative health narratives that mix useful, misleading, and dangerous content. As platforms have scaled back moderation and movements such as “Make America Healthy Again” (MAHA) have mainstreamed skepticism toward institutional expertise, the reach of health misinformation on short‑form video has become a pressing concern. Yet understanding this landscape requires more than adjudicating true versus false claims: it requires tracing how signals of credibility are performed and received.

Accordingly, we adopt context‑sensitive approaches that consider how information is structured and for what purpose \cite{Wardle_2017, Vraga_Bode_2017, Brennen_Simon_Nielsen_2021}. This aligns with work that examines the social processes through which claims are validated or contested \cite{Marres_2018, Jasanoff_Simmet_2017} and with research on performances of lay expertise and authenticity that compete with institutional authority \cite{Goffman_1959, Alexander_Giesen_Mast_2006, Au_Eyal_2022, Chun_2021}. In the contemporary wellness economy, credibility is also entangled with platform incentives, influencer branding, and commercialization \cite{Duffy_2017, O’Neill_2020, Derkatch_2018, Derkatch2023}.

Studying short‑form video at scale is methodologically challenging because meaning arises from the interplay of text, audio, and visuals \cite{Hand_2022, Laestadius_Witt_2022}. To meet this challenge, we develop a transparent, replicable annotation pipeline that combines automatic speech recognition, principled frame selection, and a single analysis model that reads both transcripts and frames, with human verification. 

We address the research question by examining the visual and rhetorical machinery that makes health advice appear credible. Focusing on nutrition and supplements, we analyze how authority signals (e.g., titles, lab coats, charts), narrative techniques (e.g., fear/urgency, references, critiques), and monetization cues (e.g., sales links, calls to action) intersect in videos on TikTok, Instagram, and YouTube.

\section{Related Work}
\label{sec:related}

\subsection{Digital health misinformation and scope}
Research documents the scale of online health misinformation \cite{Brennen_Simon_Nielsen_2021, Lundy_2023} and the influence of online personalities in shaping information flows \cite{Chen_Wang_2021}. Beyond macro accounts of declining institutional trust \cite{Waldrop_2017}, recent work urges context‑sensitive definitions that consider the availability of accurate information and the communicative purpose of content \cite{Vraga_Bode_2017, Brennen_Simon_Nielsen_2021}. Wardle’s matrix \cite{Wardle_2017} evaluates not only factual accuracy but also how content is presented (e.g., parody, manipulated media) and to what end (e.g., provocation, political advocacy, sales). We follow this pragmatic tradition by focusing on how health advice is packaged and what audiences might take as signals of trustworthiness, rather than adjudicating truth claims alone (see \cite{Chauchard_Garimella_2025}).

\subsection{Influencers and incentives}
Health and wellness culture has turned individualization into a large commercial sector and a major source of online health misinformation related to nutrition and supplements \cite{O’Neill_2020}. Influencers operate within a political economy that rewards “aspirational labor” \cite{Duffy_2017}, monetizes attention through in‑app stores and affiliate links \cite{Ohlheiser_2024}, and often positions alternative health brands against mainstream medical authority \cite{Derkatch_2018, Derkatch2023}. These dynamics intersect with political currents that valorize skepticism toward public institutions \cite{Baker_2022}. Our analysis examines how commercial cues and oppositional narratives co‑occur with scientific‑seeming packaging.

\subsection{Credibility as performance}
Credibility is context dependent, shaped by message qualities, perceived source traits, and social reputation \cite{Appelman_Sundar_2016, Ohanian_1990, Forder_2001}. Recent work extends this to online health advice and influencer culture \cite{Lederman_Fan_Smith_Chang_2014, Djafarova_Trofimenko_2019, Wellman_2023}. Building on interactionist and cultural‑sociological traditions \cite{Goffman_1959, Alexander_Giesen_Mast_2006, Reed_Alexander_2015}, we treat health advice videos as performances that mobilize lay expertise or authenticity \cite{Au_Eyal_2022, Chun_2021}. Strategies that “feel true” can privilege performed authority (titles, visuals, citations) over institutional legitimacy, a dynamic we measure empirically.

\subsection{Multimodal methods for short‑form video}
Pandemic‑era studies analyzed text and networks (e.g., Tripathi et al. \cite{Tripathi_de}; Clark et al. \cite{Clark_Lopez_Woods_Yockey_Butler_Barroso_2023}), but multimodal cues such as lab coats, certificates, charts, and production style remain underexplored in a systematic way. Studying short‑form video is challenging because meaning arises from the interplay of text, audio, and visuals \cite{Hand_2022, Miltsov_2022, Laestadius_Witt_2022}. We address this gap by systematically coding visual and rhetorical cues using a multimodal pipeline that integrates automatic speech recognition, principled frame selection, and a single analysis model that reads both transcripts and frames, with human verification. Following grounded computational social science \cite{Nelson_2020,McLevey_2021}, our pipeline enables transparent, replicable measurement of authority signals, narratives, and monetization at scale on TikTok, Instagram, and YouTube.

\section{Methodology}

\subsection{Video Corpus and Coding Scheme}
We examine how videos that disseminate questionable health information cultivate trust by analyzing a cross‑platform corpus of 152 public videos with health advice drawn from YouTube (88; 57.9\%), TikTok (48; 31.6\%), and Instagram (16; 10.5\%). Sampling proceeded in two stages. First, we seeded the corpus with content from health advice influencers named in the Center for Countering Digital Hate's 2021 report, The Disinformation Dozen. We collected a small sample from each across the three platforms (average of about three videos per influencer), with a maximum of 12 videos for one influencer and only one video for three of the named influencers. Using fresh accounts, we then primed platform recommendation systems with this content and collected additional videos via YouTube recommendations and subsequent items in Instagram Reels and TikTok feeds, retaining only videos that (a) contained some form of health advice and (b) were posted by accounts with more than 100k followers.

Each video is annotated on 26 features spanning four dimensions: (i) visual authority (e.g., lab coats, books, charts), (ii) presenter visibility and attributes (e.g., age, gender, affect), (iii) narrative strategies (e.g., urgency, references, critiques of mainstream medicine), and (iv) engagement signals (e.g., sales links, calls to action). Table \ref{tab:feature-schema} presents each feature, the guiding question, and the fixed response options to support consistent, reproducible coding.

\begin{longtable}{@{}P{0.17\textwidth} P{0.44\textwidth} P{0.31\textwidth}@{}}
\caption{Feature schema: feature name, representative question, and allowed values}
\label{tab:feature-schema}\\   
\toprule
\textbf{Feature name} & \textbf{Question} & \textbf{Allowed values}\\ 
\midrule
\endfirsthead

White lab coat & Is a white medical‑style lab coat visibly worn by anyone? & Yes / No / Unclear\\ \midrule

Medical equipment & Are medical tools (e.g., stethoscope, syringe, anatomical model) visible? & Yes / No / Unclear\\ \midrule
Background & What is the main visible setting for most of the video? & Clinic / Home / Studio or Classroom / Outdoors / Virtual / Unclear \\ \midrule

Video content type & Does a real human presenter directly appear on screen at any point, or is it only narration with no person visible? Choose “Animated” only if the entire video is a cartoon or CGI. &  Presenter‑led / Voiceover / Animated / Unclear\\ \midrule
On‑screen medical charts & Are graphical charts, anatomy images, or medical diagrams shown on screen? & Yes / No / Unclear\\ \midrule
Logos or badges & Are hospital/company/authority logos or badges visible? & Yes / No / Unclear \\ \midrule
Books or certificates & Are diplomas, certificates, or books visibly displayed? & Yes / No / Unclear\\ \midrule
Microphone/podcast setup & Are microphones or podcasting equipment (excluding clip microphone) clearly present? & Yes / No / Unclear\\ \midrule
Laboratory props & Are laboratory props (beakers, test tubes, lab equipment) visible? & Yes / No / Unclear\\ \midrule
Slides or whiteboard & Are digital slides, PowerPoint‑style visuals, or a whiteboard/chalkboard used? & Yes / No / Unclear\\ \midrule
Presenter on screen & Is a human presenter or speaker actually visible on screen (not just as a voice)? & Yes / No / Unclear\\ \midrule
Number of presenters & How many people visibly present or speak in the video? If no one is seen, use “Unclear.” & 1 / 2 / 3+ / Group / Unclear\\ \midrule
Facial expression & What is the main facial expression of the main presenter? & Confident / Scared / Neutral / Other / Unclear\\ \midrule
Age of presenter & What is the apparent age category of the main presenter(s)? & Child / Young Adult / Middle Aged / Older / Unclear\\ \midrule
Gender of presenter & What is the apparent gender of the main presenter(s)? & Male / Female / Group / Mixed / Unclear \\ \midrule
Live demonstration or experiment & Are there hands‑on demonstrations or live experiments shown by a presenter? & Yes / No / Unclear\\ \midrule
Claims a professional title & Does anyone claim or display a professional title (doctor, nurse, expert, etc.)? & Yes / No / Unclear\\ \midrule
Gives health advice & Is direct health or lifestyle advice given to the audience? & Yes / No / Unclear\\ \midrule
Fear/urgency language & Does the presenter use language designed to create fear, urgency, or alarm? & Yes / No / Unclear\\ \midrule
Personal stories/testimonials & Are personal stories or testimonials provided? & Yes / No / Unclear\\ \midrule
References to studies & Are references to published scientific studies or data provided? & Yes / No / Unclear\\ \midrule
Conspiracy mentions & Are conspiracy theories or cover‑up claims mentioned? & Yes / No / Unclear\\ \midrule
Sales pitch or links & Are products, supplements, or web links promoted or sold in the video? & Yes / No / Unclear\\ \midrule
Medical jargon & Is there significant use of technical or medical language? & Yes / No / Unclear\\ \midrule
Critiques of mainstream & Are mainstream medicine, science, or organizations criticized? & Yes / No / Unclear\\ \midrule
Call to share/subscribe & Is the viewer explicitly asked to share, subscribe, or follow? & Yes / No / Unclear

\end{longtable}

\subsection{Automated Multimodal Annotation}

\begin{figure} 
    \centering
    \includegraphics[width=0.95\linewidth]{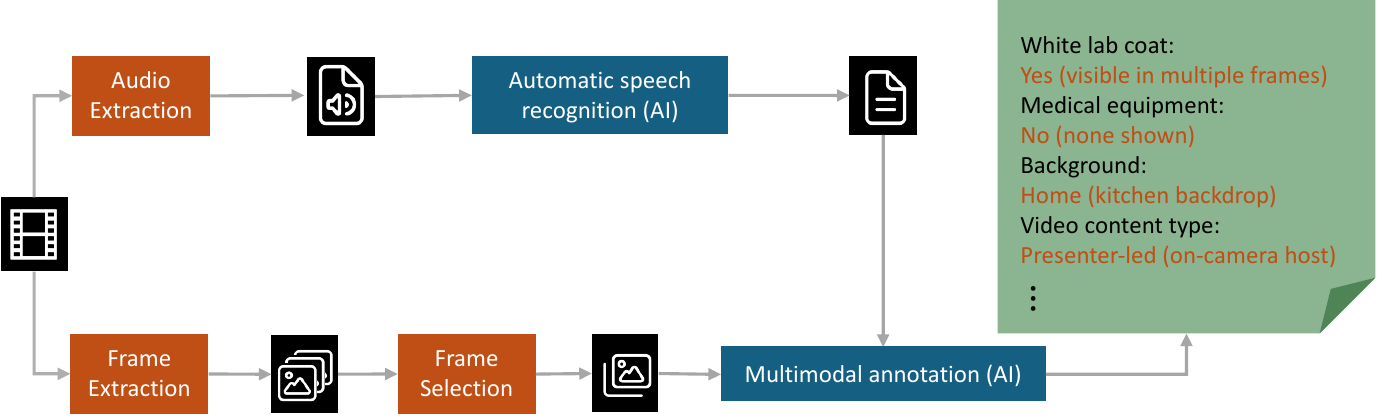}
    \caption{Automated multimodal annotation pipeline. From each video, audio is transcribed by automatic speech recognition (AI) and frames are extracted and selected; the transcript and frames then feed a multimodal annotation model (AI), which assigns values for 26 features and produces one‑sentence justifications (excerpt shown at right). Orange boxes denote scripted preprocessing, blue boxes denote AI modules, black icons indicate intermediate data artifacts, and arrows indicate processing flow.}
    \label{fig.pipeline}
\end{figure}

To generate consistent, scalable annotations of the 26 study features for every video, we designed an automated multimodal pipeline (Figure \ref{fig.pipeline}) that converts each video into two aligned inputs (a transcript of spoken content and a small set of representative frames) and then applies a single analysis model to classify all features. The pipeline is built to (i) capture both verbal claims and visual cues, (ii) avoid missing brief but meaningful visuals while limiting redundancy, (iii) produce standardized outputs suitable for quantitative analysis, and (iv) support quality assurance through human verification.

\textbf{Step 1: Speech-to-text (verbal content).} We extract each video’s audio track and produce a transcript using automatic speech recognition (ASR). This step supplies the model with the exact wording used in the video, enabling it to identify claims, references, and calls to action in the spoken narration.

\textbf{Step 2: Frame selection (visual content).} To capture visual context without overwhelming redundancy, we use two complementary sampling strategies in parallel, each producing its own frame set. The first strategy (“distinct‑frames”) extracts frames from the full video and removes near‑duplicates using a perceptual image‑similarity filter, yielding a compact set of visually unique frames. The second strategy (“uniform‑sample”) selects a small, evenly spaced sample regardless of similarity (10 frames for YouTube; 5 for TikTok/Instagram). We run the annotation model separately on each frame set and review any annotation discrepancies between these two sampling regimes to reduces the risk of missing brief but meaningful visuals while also preventing redundant frames from diluting the model’s attention.

\textbf{Step 3: Multimodal annotation.} We provide the transcript and the selected frames to a large multimodal analysis model (capable of interpreting both text and images). The model assigns a value for each of the 26 features and, for transparency, supplies a one‑sentence justification for every decision. To standardize outputs for analysis, the model fills a structured response template with fixed options for each field, recorded as feature name, chosen value, and one‑sentence justification. The instruction given to the model is summarized in Figure \ref{fig:prompt}.

\paragraph{Verification and agreement}
To evaluate the pipeline’s reliability, we drew a stratified random sample of 30 videos that preserved the platform proportions of the full corpus. These videos were manually annotated using the same 26‑feature schema. We then compared the manual labels with the pipeline’s outputs and observed strong agreement between the two.

\begin{figure}
    \centering
\begin{boxC}[\faIcon{clipboard-list}\ Prompt Template]
\textbf{You will receive:} (a) the video transcript and (b) a small set of frames.

\medskip
\textbf{Task:} For each feature in Table \ref{tab:feature-schema}, choose exactly one allowed category and provide a one‑sentence justification grounded in the transcript/frames.

\medskip
\textbf{Output:} For every feature, return the feature name, the selected value (from Table~\ref{tab:feature-schema}), and the one‑sentence justification.

\medskip
\textbf{Constraints:} Use only the categories in Table~\ref{tab:feature-schema}. Select ``Unclear'' when evidence is insufficient. Do not include any additional text.
\end{boxC}
    
    \caption{Summarized prompt template directing the multimodal annotation model to assign values for all 26 features and provide one‑sentence justifications, constrained to the categories in Table \ref{tab:feature-schema}.}
     \label{fig:prompt}
\end{figure}

\section{Results and Analysis}

We report findings from two complementary descriptive analyses of the 152 videos. First, we summarize the marginal distributions of each coded feature to characterize the typical presentation style. Second, we examine pairwise associations via cross‑tabulations to identify recurring co‑occurrence patterns (e.g., whether claiming a professional title coincides with references to studies or fear/urgency language).

Across the 152 videos, the dominant format is a confident, single presenter speaking directly to camera in a studio or classroom setting. Presenter-led videos account for 88\% of the sample, and 70\% feature just one speaker. Settings are mostly Studio/Classroom (49\%) or Home (38\%), with real clinical environments almost absent (1 video). The most common profile is a middle‑aged (55\%), male (51\%) speaker with a confident affect (83\%). Almost all videos use medical jargon (89\%) and give direct advice (99\%). About half reference studies (51\%), 43\% criticize mainstream medicine or science, 58\% include personal stories, 37\% use fear or urgency, 47\% promote products or links, and 44\% ask viewers to share or subscribe.

Authority is constructed less through authentic clinical context and more through visual and discursive cues. Lab coats are rare (5\%), but when they appear, the videos consistently claim professional status (100\%), use jargon (100\%), give advice (100\%), and almost always deploy slides (86\%), references (71\%), and calls to action (86\%); notably, these lab‑coat videos are not set in clinics but in studio or virtual spaces. Displayed books or certificates (22\%) similarly coincide with intensified persuasion: fear or urgency rises to 58\% versus 31\% when they are absent, references increase to 73\% versus 45\%, critiques of mainstream science to 67\% versus 37\%, and sales links to 73\% versus 40\%. The use of slides or whiteboards (40\%) and on‑screen medical charts (38\%) strongly co‑occur with science‑like packaging—100\% and 98\% jargon respectively—and much higher citation rates (69\% with slides, 67\% with charts), along with more critiques of mainstream (about 54\% in both cases) and more frequent calls to share or subscribe (about 61\%). Logos or badges (36\%) are disproportionately concentrated in voiceover videos: roughly 31\% of logo‑bearing videos are voiceover compared with 1\% when logos are absent, suggesting staged institutional cues stand in for a visible presenter.

Several persuasion “bundles” recur. Videos that explicitly claim a professional title (55\% of the sample) are much more likely to use fear or urgency (51\% vs 19\% when no title is claimed), to cite studies (68\% vs 29\%), to criticize mainstream medicine or science (60\% vs 22\%), to mention conspiracies (22\% vs 3\%), and to ask viewers to share or subscribe (61\% vs 25\%). Fear or urgency appears to be a central amplifier: when it is present (37\% overall), critiques of mainstream spike to 82\% versus 21\% when fear is absent, conspiracies rise to 32\% versus 2\%, references to 71\% versus 40\%, sales links to 61\% versus 39\%, calls to subscribe to 68\% versus 30\%, claims of professional title to 75\% versus 42\%, and slide use to 52\% versus 33\%. Rather than serving as safeguards, references and charts seem woven into emotive, adversarial narratives.

Personal stories work in tandem with this science‑styled persuasion. When videos include stories (58\% overall), conspiracies appear more often (21\% vs 3\% without stories), as do critiques of mainstream science (61\% vs 18\%), calls to share or subscribe (56\% vs 29\%), and references to studies (60\% vs 40\%). Nearly all story‑based videos show a visible human presenter (99\%), underscoring how testimonials function as a performance of credibility that complements citations and instructional visuals.

References to studies are not markers of rigor in this corpus; they are part of a persuasive style. When references are present (51\% overall), videos show much more fear or urgency (51\% vs 22\% without references), stronger critiques of mainstream science (62\% vs 24\%), more conspiracies (22\% vs 4\%), more sales links (56\% vs 37\%), and more CTAs (65\% vs 22\%). Jargon is universal in this group (100\%). This pattern is consistent with “sciencewashing,” where citation‑like elements and visuals give an aura of scientific credibility to emotive or oppositional messaging.

Monetization and oppositional framing co‑occur frequently. Videos with sales links (47\%) more often include conspiracies (23\% vs 5\% without sales), critiques of mainstream science (61\% vs 28\%), fear or urgency (48\% vs 27\%), and references to studies (62\% vs 42\%), and they more frequently push CTAs (61\% vs 30\%). The most rhetorically loaded subset is the cluster that mentions conspiracies (13\% overall): fear or urgency is present in 90\% of these videos, critiques of mainstream in 100\%, professional title claims in 90\%, personal stories in 90\%, references in 85\%, sales links in 80\%, and CTAs in 75\%, with most shot in studios and many using podcast‑like setups.

Specific production formats intensify these bundles. A podcast or visible microphone setup (32\% overall) concentrates multiple presenters (two speakers in 39\% vs 8\% when no microphone), a studio/classroom backdrop (84\% vs 33\%), and a male‑heavy mix (76\%). These videos show higher rates of critiques of mainstream science (57\% vs 37\%), conspiracies (22\% vs 9\%), and personal stories (71\% vs 52\%), suggesting the “expert panel” aesthetic functions as a vehicle for oppositional, narrative‑driven persuasion. Similarly, the “teacher mode” of slides or whiteboards consistently carries high jargon (100\%), more references, more fear, and more CTAs. Against this, authentic clinical context is nearly nonexistent; even when lab coats or equipment appear, they are staged in studios or virtual sets, underlining how performed authority substitutes for lived clinical authority.

Demographic patterns are descriptive rather than causal but notable. Male presenters (51\%) are more often in studio/podcast formats, with higher rates of references (60\%), critiques of mainstream science (50\%), and CTAs (55\%), while female‑presenter videos (29\%) are more home‑based (59\%) and include more live demonstrations (23\% vs 10\% for males), with somewhat lower conspiracies (11\%) and CTAs (32\%). Middle‑aged presenters (55\%) dominate the high‑intensity persuasion profiles(fear or urgency, 43\%, and critiques of mainstream, 57\%), whereas young adults (22\%) show lower levels of fear (18\%) and critiques (27\%). Confidence is the prevailing affect across groups, especially in presenter‑led content (94\% “Confident”).

Taken together, the corpus suggests a consistent packaging for health misinformation: a studio/classroom “edutainment” style where performed authority (titles, slides, charts, jargon, and citations) is fused with emotional activation (fear and urgency), oppositional narratives (critiques of mainstream science and conspiratorial frames), and commercialization (sales links and CTAs). The same ingredients appear to travel together, and it is this bundle, rather than any single marker, that best characterizes the persuasive style.

\section{Contributions and Significance}

Our contributions are threefold:

\textbf{Code:} We propose a multimodal annotation pipeline that combines automatic speech recognition, complementary frame selection, and schema‑constrained labeling with human verification. The pipeline enables transparent, reproducible measurement of visual and rhetorical cues in short‑form video, a medium that is otherwise difficult to analyze at scale.

\textbf{Dataset:} We provide an annotated corpus of 152 health advice videos, labeled on 26 features spanning visual authority, presenter attributes, narrative strategies, and engagement/monetization signals. The cross‑platform focus (TikTok, Instagram, YouTube) and nutrition/supplement domain make the corpus useful for comparative research, benchmarking, and mixed‑methods studies.

\textbf{Findings:} We show that visual authority cues scaffold rhetorical and commercial strategies in health misinformation, yielding recurrent bundles that fuse performed credibility (titles, slides, charts), emotional activation (fear/urgency, oppositional framing), and monetization (sales links, calls to action). This pattern clarifies how “science‑styled” packaging can co‑occur with adversarial narratives, refining accounts of credibility in multimodal misinformation.

\section{Conclusion}
This study investigates how authority signals, narrative techniques, and monetization strategies intersect in health advice videos, with a focus on nutrition and supplements. Across 152 TikTok, Instagram, and YouTube videos, we find a dominant studio or home edutainment style with confident single presenters and few clinical settings. Credibility is performed through recurring bundles that pair titles, charts, slides, and citations with jargon, fear or urgency, critiques or conspiracies, and sales links or calls to subscribe; references and science like visuals often scaffold emotive, oppositional narratives. This advances a social account of credibility as performance and offers a transparent multimodal annotation pipeline and dataset. Limitations include a modest, cross sectional, English corpus and automated labeling.

\bibliographystyle{unsrt}  
\bibliography{references}  



\end{document}